\documentclass[useAMS,usenatbib]{mn2e}

\usepackage{times}

\usepackage{graphicx}
\newcommand{\avg}[1]{\left< #1 \right>}

\title[Accretion onto dark matter halos and subhalos]
{The environment and redshift dependence of accretion onto dark matter halos and subhalos}
\author[H. Tillson, L. Miller \& J. Devriendt]{H. Tillson$^{1}$\thanks{email: Henry.Tillson@astro.ox.ac.uk}, 
L. Miller$^{1}$ \& J. Devriendt$^{1,2}$\\
$^1$Department of Physics, University of Oxford, The Denys Wilkinson Building, Keble Road, Oxford, 
OX1 3RH, UK\\
$^2$Centre de Recherche Astrophysique de Lyon, UMR 5574, 9 Avenue Charles Andr$\acute{e}$, 
F69561 Saint Genis Laval, France}
\begin{document}

\date{Accepted 2011 June 22.  Received 2011 June 22; in original form 2010 September 27}

\pagerange{\pageref{firstpage}--\pageref{lastpage}} \pubyear{2011}

\maketitle

\label{firstpage}

\begin{abstract}
A dark-matter-only Horizon Project simulation is used
to investigate the environment- and redshift- dependence of
accretion onto both halos and subhalos. 
These objects grow in the simulation via mergers 
and via accretion of diffuse non-halo material, 
and we measure the combined signal
from these two modes of accretion. 
It is found that the halo accretion rate varies less strongly with redshift than
predicted by the Extended Press-Schechter (EPS) formalism and
is dominated by minor-merger 
and diffuse accretion events at
$z=0$, for all halos. These latter growth mechanisms
may be able to drive the radio-mode feedback hypothesised for recent galaxy-formation
models, and have both the correct accretion rate and form of cosmological evolution.
The low redshift subhalo accretors in the simulation
form a mass-selected subsample safely above the mass resolution limit
that reside in the outer regions of their host,
with $\sim 70\%$ beyond their host's virial radius,
where they are probably not being significantly stripped of mass.
These subhalos accrete, on average, at higher rates than halos at low redshift 
and we argue that this is due to their enhanced clustering at small scales.
At cluster scales, the mass accretion rate onto halos and subhalos
at low redshift is found to be only weakly dependent on environment 
and we confirm that at $z\sim2$ halos accrete independently of 
their environment at all scales, as reported by other authors.
By comparing our results with an observational study of black hole growth, we
support previous suggestions that at $z>1$, dark matter halos and their
associated central black holes grew coevally, but show that by the
present day, dark matter halos could be accreting at fractional rates that
are up to a factor $3-4$ higher than their associated black holes. 
\end{abstract}


\section{Introduction}

In the $\Lambda$CDM model, structures are seeded with initial fluctuations and merge to form bound, 
virialized dark matter halos that become more massive as the universe ages. Luminous galaxies form as 
baryonic matter cools and condenses at halo centres \citep{White_Rees,Fall,Blumenthal}. Dense dark halos, 
however, often contain embedded subhalos and it has been demonstrated that low mass subhalos can survive 
in their hosts for several billion years \citep{Tormen_97,Tormen_98,Moore}. One challenge for cosmological 
N-body simulations is to link dark matter halos and subhalos with luminous galaxies \citep{Bower06,Conroy_06,Vale}. 
Understanding this relationship has proved difficult \citep{Diemand04, Gao04, Nagai} and most explanations are 
provided by semi-analytic models 
(\citealt{White&Frenk1991,Somerville&Primack,Hatton03,Bower06,Cattaneo06}; \citealt{Croton06}, hereafter C06). 
Nonetheless, a vital ingredient in explaining luminous galaxy growth in large groups and 
clusters is an understanding of how dark matter halos and subhalos accrete mass in dense environments. 

The standard implementation of the Extended Press-Schechter (hereafter EPS) 
formalism \citep{Bond, Lacey} can be used to analytically compute the average mass 
accretion onto a halo of mass $M_{H}$. \cite{Miller06} (hereafter M06) showed that:

\begin{equation}\label{eMill}
\avg{\dot{M}_{H}} \simeq M_{H}\left|\frac{d\delta_{c}}{dt}\right|f(M_{H}),
\end{equation}

\begin{equation}\label{eMill2}
\frac{d\delta_{c}}{dt}=\frac{d\delta_{c}}{dD}\frac{dD}{dz}\frac{dz}{dt}
\end{equation}
where 
$\delta_{c}(t)$ is the critical density contrast above which an object will collapse to form 
a bound structure, $D(z)$ is the linear growth factor and 
$f(M_{H})$ is a weak function of halo mass (for alternative analytic 
expressions for halo growth derived using EPS theory, see \citealt{Hiotelis06} and \citealt{Neistein08}). 
Equation (\ref{eMill}) can in principle be used for all redshifts and halo masses, 
but a recent simulation study by \cite{Cohn08} tested it against 
the accretion histories of massive halos at $z=10$ and found that it overestimated 
their accretion rate. 

A simplified assumption of the EPS framework inherent in equation (\ref{eMill}) is that 
halos accrete at rates that do not depend on their environment. 
This restrictive assumption, however, is not a prediction of the theory and so 
various authors have recently relaxed it. 
\cite{Sandvik} implemented a multidimensional generalization of
the EPS formalism and used an ellipsoidal collapse model where collapse depended 
both on the overdensity and the shape of the initial density field.
They found only a weak dependence between halo formation redshift and halo
clustering which was stronger for more massive halos,
in disagreement with the reported halo assembly bias in numerical
simulations \citep{Gao05, Gao07, Maultbetsch}. 
\cite{Zentner} modified the EPS formalism by using a Gaussian smoothing window function,
and \cite{Desjacques} allowed the density threshold to have an environment dependence, but
both authors found that dense large-scale environments preferentially 
contain halos that form later. 
We are hence lacking an EPS model that is able to account for halo assembly bias
and predict a modified analytic version of equation (\ref{eMill}) for
the halo accretion rate.
Deviations from the EPS accretion rate are therefore expected 
in the highly non-linear regime of cluster formation at $z<1$, as equation (\ref{eMill}) 
cannot account for accretion onto subhalos embedded within larger halos.

To date, several authors have defined prescriptions for computing accretion onto halos 
using dark-matter-only simulations:

\begin{itemize}
\item \cite{Wechsler} $-$ henceforth W02 $-$ identified the mass accretion history (hereafter MAH) 
of $\sim14000$ halos at $z=0$ using the ART code \citep{Kravtsov} in a WMAP1 cosmology. 
Using their algorithm, W02 found that the accretion histories of their 
present day halos were, on average, well fitted by:

\begin{equation}\label{eWech}
M_{H}(z)=M_{0}e^{-\alpha(z_{f})z}
\end{equation}
where 
$M_{0}$ is the present day mass of a halo and 
$\alpha(z_f)$ is a parameter which describes its formation epoch. 
Ignoring the slight 
mass dependencies of $\alpha(z_{f}(M_{H}))$ and the $f(M_{H})$ term in equation (\ref{eMill}), 
it can be seen that equation (\ref{eWech}) is a sensible fit for W02 to have chosen because 
in the case of an Einstein-de-Sitter (EdS) universe, their $\avg{\dot{M}_{H}}$ has 
the same $M_{H}\dot{z}$ dependence as equation (\ref{eMill}), differing only in normalization 
($d\delta_{c}/dz=1.686$ for an EdS universe).

\item \cite{VanDenBosch} used the N-branch merger tree algorithm of \cite{Somerville} and 
found that a two parameter fit better described the MAHs of his halos, although 
M06 demonstrated that this two parameter fit becomes unphysical locally as it predicts that 
present day halos are not accreting mass. \cite{VanDenBosch} also provided a relation for 
$\alpha$ and $z_{f}$ that can be used in equation (\ref{eWech}):

\begin{equation}\label{eBosch}
\alpha=\left(\frac{z_{f}}{1.43}\right)^{-1.05}
\end{equation}
but it is more common to define $z_f$ as the epoch at which the present day halo of 
interest had half of its present day mass: 

\begin{equation}\label{eAlpha}
z_{f}=\frac{\ln{2}}{\alpha}
\end{equation}

\item More recently, \cite{McBride} investigated the MAHs of $\sim500000$ halos from the Millennium simulation 
with $M_{H}>10^{12}M_{\odot}$ and $0\leq z\leq 6$ and found that only $\sim 25\%$ 
were well described by equation (\ref{eWech}). 
They introduced a second parameter, $\beta$, and showed that a function of the form:

\begin{equation}\label{eMcBride}
M_{H}(z)\propto (1+z)^{\beta}e^{-\gamma z}
\end{equation}
provided a better fit to the halo MAHs. 
\item \cite{Fakhouri10} used a joint dataset from the Millennium I and II simulations 
and found that equation (\ref{eMcBride}) held across five decades in mass up to $z=15$.
\end{itemize}

These listed accretion fits only apply when averaged across all environments. 
In order to understand accretion in dense regions such as clusters, one must resolve 
substructure and design an accretion algorithm that can account for accretion onto halos
and all levels of substructure. 
The difficulties in devising such an accretion algorithm are two-fold: firstly, 
it should define a single progenitor for each and every (sub)halo which 
accurately represents that object at earlier epochs, and secondly, it 
must conserve mass (which becomes harder to do when one introduces subhalos). 
In this study, outputs from a high resolution dark-matter-only N-body simulation 
have been used and a new robust method for defining accretion onto halos and subhalos is provided, 
building on previous simulation studies and moving beyond EPS theory. 
The primary aim is to investigate exactly how accretion onto halos and subhalos 
behaves as a function of redshift, mass and environment.

One way of measuring a (sub)halo's environment is to compute the two-point correlation function, 
as this yields information on halo bias or degree of clustering. 
\cite{Percival} used four $\Lambda$CDM simulations with differing box sizes, 
$\sigma_{8}$ values and particle masses, with each simulation containing $256^3$ particles, 
to examine four different halo merger samples at $z=2$. They found no difference in 
clustering at this redshift between the merger samples of halos of a given mass. 
We examine the clustering of halos and subhalos in a higher resolution simulation and 
test this conclusion at $z\sim2$ and at lower redshifts.

A natural corollary is then to investigate whether the dark matter distribution alone has any relevance to 
SFR/galaxy downsizing \citep{Cowie,Brinchmann,Bauer,Bundy,Faber,Panter} and AGN downsizing 
\citep{C03, Steffen, Barger, Hasinger, Hopkins}. AGN feedback provides a 
plausible explanation of galaxy downsizing, and has been successfully 
implemented in semi-analytic models (\citealt{Bower06, Cattaneo06}; C06) and has been 
observed as the phenomenon responsible for the suppression of star formation 
in ellipticals in the local universe \citep{Schawinski07, Schawinski09}. 
AGN downsizing is less well understood and is a two-fold degenerate phenomenon 
driven either by low mass black holes accreting at near-Eddington rates 
\citep{Heckman} or by supermassive black holes accreting at low rates \citep{Babic}. 

The structure of this paper is as follows. 
Section \ref{simulation_section} describes the N-body simulation that was used and 
Section \ref{describe_algs_section} explains the accretion algorithm. 
Section \ref{results_section} examines accretion onto halos and subhalos within groups and clusters and draws comparisons with EPS and W02. 
Section \ref{discussion_section} discusses the implications of the results of this paper and
Section \ref{conclusions_section}, the final section, lists our conclusions. 
A WMAP3 cosmology has been adopted throughout with $\Omega_m=0.24, \Omega_\Lambda=0.76, \Omega_b=0.042, n=0.958, h=0.73$ 
and $\sigma_8=0.77$. All masses are in units of $M_{\odot}$.

\section{The simulation}\label{simulation_section}

We have analyzed outputs from one of the Horizon Project simulations\footnote{http://www.projet-horizon.fr} 
which used the \verb!GADGET-2! code \citep{Springel05} and tracked the evolution of $512^3$ 
dark matter particles within a box of comoving side length $100h^{-1}$Mpc in a $\Lambda$CDM universe.  

The AdaptaHOP halo-finder \citep{Aubert} $-$ hereafter AHOP $-$ was used to detect halos. 
AHOP assigns a local density estimate to each particle computed using the standard SPH kernel 
\citep{Monaghan} which
weights the mass contributions from the $N$ closest neighbouring particles ($N$ is usually taken to be $20$).
Halos are then resolved by imposing a density threshold criterion and by measuring local
density gradients.
AHOP is an alternative to the popular friend-of-friend (FOF) halo-finder \citep{Davis}, which groups together 
particles that are spatially separated by a distance that is
less than typically $20\%$ of the mean inter-particle separation. 
Recently it has been demonstrated that inappropriate definitions of halo mass can
introduce large uncertainties in the halo merger rate \citep{Hopkins10} $-$ FOF, in particular, significantly overestimates
the halo merger rate for halos that are about to merge \citep{Genel09}, and so we avoid using it.
For a critical quantative comparison between AHOP and FOF, see \cite{Tweed}.

In order to detect substructure we have used the Most massive Sub-node 
Method \citep{Tweed} $-$ hereafter MSM $-$ which successively raises the density 
thresholds on the AHOP halo until all of its node structure has been resolved. 
The most massive leaf is then collapsed along the node tree structure to define a main halo,
and the same process is repeated for the lower mass leaves, defining substructures 
of the main halo. 
For detailed descriptions of alternative subhalo-finders, like \verb!SUBFIND!, see \cite{Giocoli}. 

The output timesteps from the Horizon simulation were separated by
$0.01$ in scale factor from $z=99$ to the present day, 
but we restricted our analysis to halos and subhalos 
in the redshift range $0\leq z\leq9$.
The mass of each particle, $M_p$, was $6.8 \times 10^{8} M_{\odot}$ and 
halos and subhalos with a recorded accretion value contained at least $40$ particles. 
The mass of a (sub)halo used in this study corresponded to the total mass, $M_{T}$, 
detected by the halo-finder.
For reference, the MSM algorithm resolved 223781 objects at $z=0$ and $\sim20\%$ of these objects were subhalos. 
The TreeMaker code \citep{Tweed} was then used to link together all the time outputs by 
finding the fathers and sons of every halo and subhalo.

\section{Devising a halo and subhalo accretion algorithm}\label{describe_algs_section}

This section comes in three main parts. We begin by defining the main branch onto a given object 
(``object'' henceforth refers to halos and/or subhalos). We then provide an algorithm which 
identifies objects that take part in fake mergers. 
The section concludes with an explanation of the algorithm that was used to compute accretion onto bound halos and subhalos. 

\subsection{A simple merger}

In Fig.\ref{simple_merge-schematic}, halo $i$ and $j$ at timestep $t_{2}$ merge to form halo 
$k$ and $l$ at timestep $t_{1}$, where $t_{1} > t_{2}$. In order to compute the 
accretion rate onto $k$, one must define a `main father' for $k$ and various authors have 
adopted different prescriptions for identifying the main father of a halo (\citealt{Springel01}; W02). 
W02, for example, define the main father of $k$ as the halo that contributes the most 
mass to $k$ but require the main father's most bound particle to be part of $k$ if the main father 
is not at least half $k$'s mass. These rules force each halo to have a single main son and a single main father. 

There is freedom to choose the main father of $k$ as either the physically most massive father or the 
father that contributes the most mass. We have found little difference between results obtained 
from using these two definitions and so we adopt the latter definition throughout. 
In Fig.\ref{simple_merge-schematic}, halo $k$'s main father is $j$ and the main branch is shown by the solid line. 

\subsection{Anomalous events}

Anomalous events describe halos that spatially coincide at one timestep and then separate at later timesteps. 
These halos might take several timesteps to form a bound merger halo or they might never coincide again. 
One must hence be careful that their accretion estimator accounts for accretion onto bound objects only. 

\begin{figure}
\includegraphics[width=8cm,height=4cm]{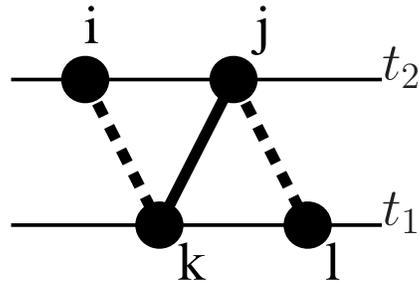}
\vskip-1.5em
\caption{Halo $i$ and $j$ at timestep $t _{2}$ merge and form two halos, $k$ and $l$ at the later 
timestep $t _{1}$. Halo $k$'s main father is $j$ and the main branch is shown by the solid line.}
\label{simple_merge-schematic}
\end{figure}

To illustrate this point further, one would na\"{\i}vely expect 
the mass accretion rate of halo $k$ in Fig.\ref{simple_merge-schematic} 
at timestep $t_1$ to be $(M_{k}-M_{j})/(t_1-t_2)$ but when this is applied to all the halos at timestep $t_1$ 
there are a larger than expected number of negative accretion events 
(halos aren't losing mass in the hierarchical halo growth paradigm). 
Physically it is perfectly possible for mergers to result in mass loss 
along the main branch, as during a merger process, material is stripped from bound objects. 
A system of objects undergoing a merger will, however, eventually form relaxed, bound 
objects at later times and so pinpointing the time interval during which mass is accreted is crucial
(we do not measure mass loss via stripping in this work).

\subsubsection{Identifying anomalous events}\label{anomalous_section}

Testing to see whether an object is bound is one definitive way of excluding such fake 
events and it is common practise to sum the kinetic and potential energies of each object 
and disregard those objects whose total energy is positive \citep{Maciejewski}. 
We combine this technique with an independent anomalous detection method to identify
unbound objects at each redshift.

Our prescription for identifying objects participating in anomalous events is as follows. 
The fathers of an object $k$ at timestep $t_2$ are found and if object $k$ has two or more 
halo fathers that each donate a mass $M_{D}\geq 20M_{p}$, then object $k$ is flagged as a 
possible fake merger candidate. ($20M_{p}$ is chosen here rather than the mass resolution 
limit of $40M_{p}$ used in later sections, because $20M_{p}$ is a common mass resolution 
limit used in other simulation studies and it also maximises the number of possible anomalous events.) 
The sons of $k$ are then found and if $k$ donates a mass $M_{D}\geq 20M_p$ to two or more halos, 
then it has fragmented and it is identified as an anomalous event candidate. In the case of AHOP 
halos in this study, which average over their environment and whose substructure is not resolved, this 
is the sole anomalous event criterion and the same criterion is then imposed on the next halo at timestep $t_2$. 

For subhalos an additional condition is imposed. 
Imagine that two halos at timestep $t_{3}$ merge to form a halo, $k$, which hosts a subhalo 
at the subsequent timestep $t_{2}$. Halo $k$ and its subhalo are then detected
as separate halos at the following timestep $t_{1}$ ($t_{1}>t_{2}>t_{3}$). 
This system has transitioned over three timesteps from two halos, 
to a halo and a subhalo and back to two halos again, and is hence an anomalous event as no merger has taken place. 
The subhalos of a given host halo are therefore also examined if the host does not fragment.
If a subhalo at $t_{2}$ donates a mass $M_{D}\geq 20M_{p}$ to a halo at 
$t_{1}$ that is a different halo to the halo son of its host,
then it is identified as part of an anomalous event, as are its subhalos (if it has any) and 
its host. 
The key ideas of this anomalous event detection method are therefore:

\begin{itemize}
\item searching for channels that receive/donate at least $20M_{p}$ from/to two or more different halos and 
\item ensuring that the host and all associated substructures are flagged in the case of 
any one of these objects being classified as participating in an anomalous event. 
\end{itemize}

\subsubsection{Identifying unbound objects}\label{no_anom_section}

Table \ref{Tanom} assesses the relative importance of unbound MSM objects above the mass 
threshold in the simulation ($M\geq 40M_{p}$) for each of the redshifts shown in column 
$1$ (these redshifts have been chosen because the number of subhalos increases with 
decreasing redshift in the simulation, as clusters form). The percentages in Table \ref{Tanom} 
express the number of objects above the threshold mass satisfying the condition in 
each column as a fraction of the total number of objects above the threshold mass 
at the redshift in question, with the exception of the bracketed values in column $3$, 
which show the fraction of anomalous events that are unbound. 

There is a positive correlation between the independently identified anomalous events and unbound objects, 
with a large fraction of the anomalous events being unbound (henceforth unbound refers either to an 
object with total energy $E_{T}\geq0$ or an object participating in an anomalous event or both). 
Not all objects in column $3$ have $E_{T}\geq0$, however, and so there is a small 
population of unbound objects at each redshift that would be missed if just a 
requirement of $E_{T}\geq0$ were imposed on every object. 

\begin{table}
  \centering
  \caption{The relative importance of unbound MSM halos and subhalos above the threshold mass ($M\geq40M_{p}$) 
in the $512^{3}$ simulation.}
 \resizebox{\columnwidth}{1.25cm}
           {
             \begin{tabular}{@{}ccccc@{}}
               \hline
               Redshift & $E_{T}\geq 0$ & Anomalous ($E_{T}\geq0$) & Objects with \\
               &$\%$&$\%$&recorded accretion\\
               & & & $\%$\\
               \hline
               $0.49$ & 23.1 & 7.85 (84.8) & 73.5 \\
               $0.23$ & 23.7 & 8.57 (87.3) & 73.2 \\
               $0.01$ & 24.1 & 9.14 (89.7) & 73.4 \\
               \hline
             \end{tabular}
           }
           \label{Tanom}
\end{table}

Only bound objects above the mass threshold can have a recorded accretion value in this study, 
despite $\sim 38\%$ of all the objects at each of the redshifts shown in Table \ref{Tanom} having a mass 
below the chosen threshold limit. Bound objects below threshold, however, are not removed from the sample 
and so it is possible for a bound object with $M<40M_{p}$ to be a main father. 
We therefore avoid biasing the accretion events in the simulation, whilst 
ensuring that only well resolved objects have an accretion value. 

Column $4$ shows the fraction of objects above the mass threshold with a recorded accretion value. 
A very small fraction of bound objects with $M\geq 40M_{p}$ do not have a measured accretion rate because 
they do not satisfy some additional criteria imposed by the accretion algorithm, which we explain in the following section.

\subsection{The accretion algorithm}\label{shalo-accn-alg-section}

In detecting substructure, \cite{Springel01} required that several of the most bound particles of the main 
father were included in the main son $-$ this was more robust than tracking the evolution 
of the single most bound particle, which essentially performs a random walk across time. 
We have defined the main son as the son which 
receives the most mass from the object of interest, 
consistent with our main father definition. 

\begin{figure}
\includegraphics[width=8cm,height=4cm]{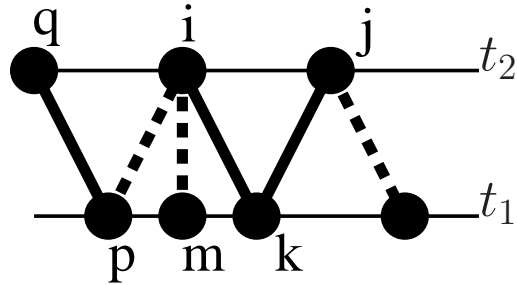}
\vskip -1em
\caption{A schematic illustrating the halosub accretion algorithm that accounts for accretion onto halos and subhalos. 
In this example object $k$, whose main father is object $j$ (solid line), 
has been identified as the main son of object $i$ (solid line). 
The accretion onto $k$ using the halosub method is therefore $(1-f_{j})M_{k}$,  
where $f_{j}$ is the fraction of $k$'s mass that comes from $j$.
Object $m$ is not the main son of object $i$ and because it doesn't have any other fathers it is skipped. 
Object $p$'s main father is $q$, hence the accretion onto $p$ is $(1-f_{q})M_{p}$. 
The halosub method therefore only ever records zero or positive accretion rates.}
\label{halosub-accn-schematic}
\end{figure}

We shall henceforth refer to the algorithm that computes accretion onto halos and associated substructures 
as the ``halosub'' method and it is illustrated in Fig.\ref{halosub-accn-schematic}. 
For object $i$ at timestep $t_{2}$ the main son $k$ (solid line) is identified. 
Using our main son definition this means that most of $i$'s mass goes to $k$ and the 
remainder goes to $m$ and $p$. The father that contributes the most mass to $k$ is then found; 
in this example $j$ is the main father (solid line). The mass accretion onto $k$ is 
therefore $(1-f_{j})M_{k}$ where $f_{j}$ is the fraction of $k$'s mass that comes from 
object $j$. Object $k$ is now flagged and the accretion onto the other sons of $i$, $m$ and $p$, is considered. 
Since $m$ is not the main son of $i$ and $m$ doesn't have any other fathers, an accretion value for 
$m$ is not recorded and it is flagged as an orphan. If however one of the sons, $p$, of the object of 
interest does experience mass accretion, we identify the main father, $q$, and record the mass accreted: 
$(1-f_{q})M_{p}$. Object $p$ would then also be flagged. To summarise, 
we list the principal features of the halosub method:

\begin{itemize}
\item the measured mass accretion onto an object represents the sum of 
diffuse accretion (material not bound to any resolved structure) and merger-driven growth
\item mass loss events are considered to be zero accretion events:
measured accretion signals in this study are never negative
\item all objects with a recorded accretion value are bound and have a mass $M\geq 40M_{p}$
\item no distinction is made between halos and different levels of substructure
\end{itemize}

Since we do not attempt to measure the mass lost from an object during a given time interval,
the accretion rate in this study can be thought of as an upper limit.
Note that objects which only lose mass and have a recorded accretion rate of zero
are identified as systems where the bound main son of the object of interest has only one bound father.
A flagged object means that either the accretion onto that object has already been accounted for 
or that object has been identified as an orphan. 

\subsection{Limitations}

Other than finite mass and time resolutions which are shortcomings of any simulation, 
we consider the growth of halos and subhalos in a $\Lambda$CDM universe without 
a prescription for the gas physics. The dark-matter-only simulation satisfies the objective of this study, 
however: to determine whether halo and subhalo accretion is dependent on environment. 
The accretion algorithm excludes tidal stripping from the measured accretion rate but objects are 
stripped of mass in the simulation as they undergo mergers and this reduces their mass.

\section{Results}\label{results_section}

Throughout this section:

\begin{enumerate}
\item ``object'' refers to halos and/or subhalos.
\item the mass of an object corresponds to the total mass, $M_{T}$, detected by the halo-finder.
\item only bound objects above the mass threshold, $M\geq 40M_{p}$, can have a recorded accretion value.
\item the measured mass accretion is the sum of diffuse- and merger-driven accretion:
we have not measured mass loss.
\item $\mu \equiv \dot{M}/M$ denotes the specific accretion rate, with units of Gyr$^{-1}$, onto an object of mass $M$.
\item $\delta \equiv \delta M_{H}/M_{H}$, where $M_{H}$ represents the mass of a halo.
\end{enumerate}

\subsection{Accretion onto dark matter halos}\label{dark_halo_accretion_section}

\subsubsection{Comparison with EPS}

\begin{figure}
\centering
\includegraphics[width=\columnwidth,height=7cm]{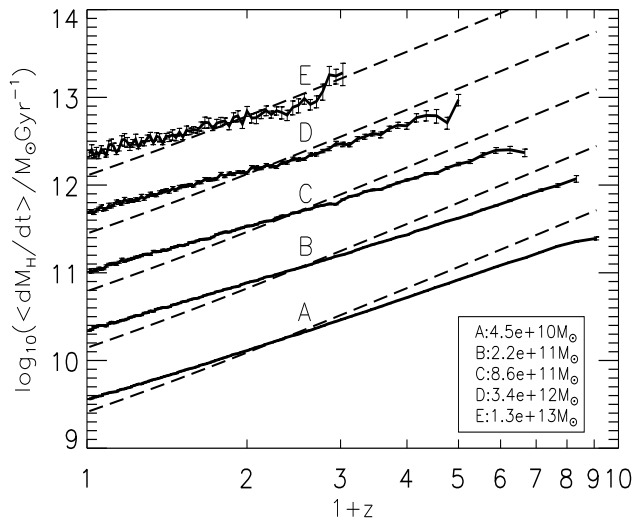}
\caption{The average halo accretion rate as a function of redshift and halo mass.
The accretion values onto the AHOP halos are shown 
as the solid lines for each of the five $\avg{M_{H}}$ bins,
with the errors corresponding to the $1\sigma$ errors on the mean accretion rate. 
The EPS curves using equation (\ref{eMill}) are shown as the dashed lines for each mass bin.}
\label{halo_accn}
\end{figure}

Fig.\ref{halo_accn} shows the average accretion rate onto the AHOP halos from 
the simulation as a function of redshift and halo mass. Halos with recorded 
accretion values are binned in mass at each redshift and the average accretion rate for 
each mass bin is computed. Averages of the corresponding mass bins over redshift then yield 
constant $\avg{M_{H}}$ values (W02 adopt an alternative technique, however, by binning 
the $z=0$ halos in mass and then averaging over all the accretion trajectories in each 
bin at each redshift). The solid lines show the accretion rates onto the AHOP halos using 
the halosub method, and the error bars indicate the $1\sigma$ errors on the mean accretion rate. 
The EPS predictions for each of the $\avg{M_{H}}$ bins, computed using 
equation (\ref{eMill}), are shown as the dashed lines.

Fig.\ref{halo_accn} shows that the simulation mass trajectories have a lower gradient 
across redshift than the EPS curves, which overestimate the accretion rate onto the 
lowest mass halos in the simulation at high redshift by a factor of $\sim 2$, and 
underestimate it by a factor of $\sim1.6-1.8$ at $z=0$. 
It is tempting to think that the enhanced accretion 
onto halos with respect to EPS theory at low redshift 
results from the exclusion of mass loss in our measured halo accretion rate.
However, EPS doesn't account for mass loss from halos either: 
halos only grow with time by construction. The offset with EPS should therefore
be regarded as an offset in gradient and Fig.\ref{halo_accn} implies
that the \cite{Lacey} EPS formalism may only require minor adjustment
to reproduce the simulated trajectories.
%We reserve a detailed discussion of the disagreement with EPS for Section \ref{EPS_section}.

\subsubsection{The different accretion modes}

\begin{figure}
\centering
\includegraphics[width=\columnwidth,height=7cm]{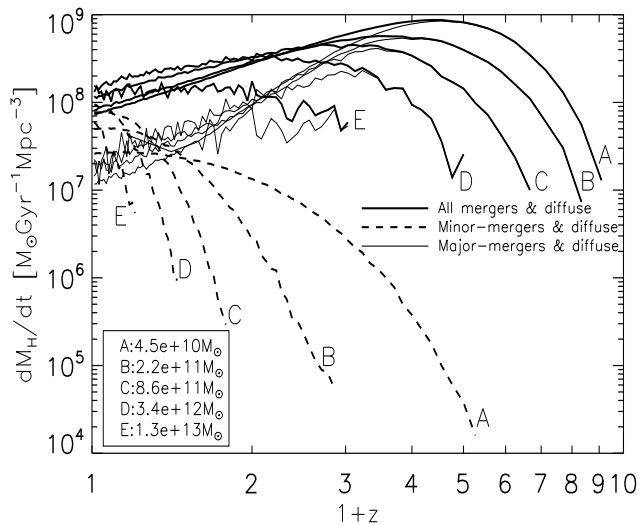}
\caption{The total mass accretion rate onto the AHOP halos per 
comoving cubic Mpc as a function of redshift, halo 
mass and $\delta$ ($\equiv\delta M_{H}/M_{H}$). 
The mass bins correspond to the $\avg{M_{H}}$ bins in Fig.\ref{halo_accn}, with 
the lower mass curves shifting to higher $z$. 
The dashed and thin solid lines show each halo mass bin decomposed into 
halos with $\delta\leq0.02$ (minor-merger $\&$ diffuse accretion) 
and $\delta\geq0.08$ (major-merger $\&$ diffuse accretion) respectively. 
The thick solid lines show the mass trajectories integrated over all $\delta$.}
\label{dm_by_m}
\end{figure}

\begin{figure*}
\centering
\includegraphics{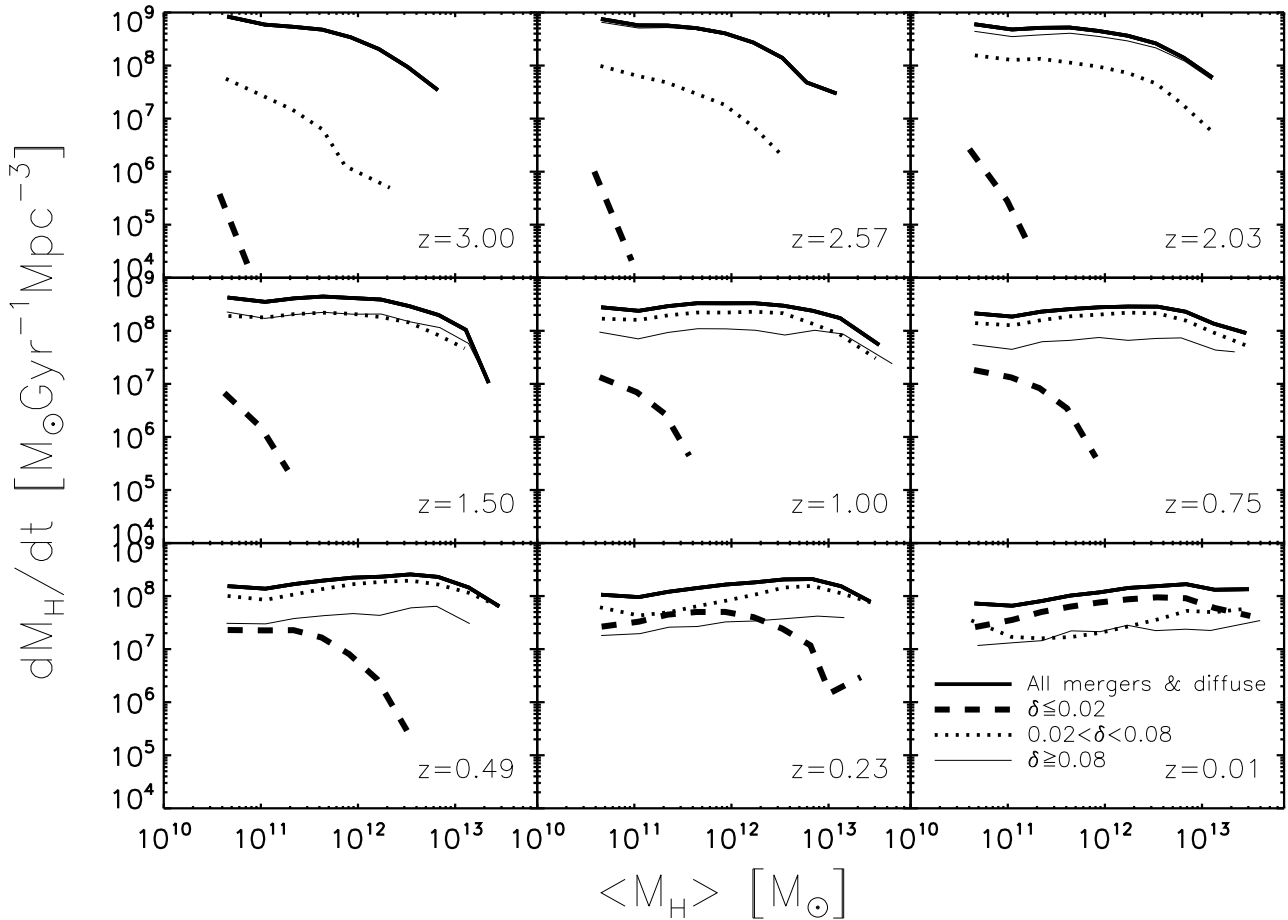}
\caption{The total mass accretion rate onto the AHOP halos per comoving cubic Mpc 
as a function of halo mass and accretion mode (denoted by $\delta$), 
shown for several redshifts. 
The linestyles have the same meaning as 
in Fig.\ref{dm_by_m} except we also include halos with $0.02<\delta<0.08$
shown by the dotted lines.}
\label{mass_func}
\end{figure*}

The mass accreted onto the AHOP halos in Fig.\ref{halo_accn} is the summed contribution of 
diffuse accretion events and minor and major-merger events, 
hence in Fig.\ref{dm_by_m} we examine the relative 
importance of these accretion modes as a function of halo mass and redshift. 
At each redshift, the dimensionless quantity $\delta$ ($\equiv \delta M_{H}/M_{H}$) 
was computed for each accretion event: 
the dashed lines and the thin solid lines show halos with
$\delta\leq 0.02$ (minor-mergers $\&$ diffuse accretion) 
and $\delta\geq 0.08$ (major-mergers $\&$ diffuse accretion) respectively. 
The total mass accretion rate per comoving cubic Mpc for halos in a given mass bin
and of a given $\delta$ at each redshift was then computed. The thick solid lines show the total mass 
accretion rate per comoving cubic Mpc integrated over all $\delta$. 
For a given linestyle, the lower mass curves shift to higher redshifts.

At high redshift, all halos are found to accrete mass 
diffusely in high fractional events with the peak in 
activity shifting to lower redshifts for more massive halos. 
As the mass accreted onto the lowest mass halos via minor-mergers and diffuse
accretion starts to plateau at low redshift, 
minor-merger and diffuse accretion activity onto the more massive halos starts to rapidly accelerate: 
low mass halos and non-halo material are being accreted onto larger structures. 
By $z=0$, the combined minor-merger and diffuse accretion signals dominate the growth of all halos. 
We further remark that the dashed curves have a similar cosmological evolution to the ``radio-mode'' 
integrated black hole accretion rate density curves found by C06 and \cite{Bower06}, but
leave a more detailed discussion for Section \ref{Croton_section}.

We have tested the ability of the cut-in-delta method 
at distinguishing between merger type 
by adopting the more classical progenitor mass ratio. Each progenitor $j$
of accretor $k$ was assumed to merge in turn with $k$'s main father $i$, 
with progenitor mass ratio $\chi\equiv M_{i}/M_{j}$, donating
$f_{j}M$ to accretor $k$ at the following timestep, 
where $f_{j}$ denotes the fraction of $k$'s mass that comes from $j$.
Events with $\chi\leq3$ ($\chi>3$) were recorded as major (minor) mergers.
We found that major mergers and diffuse accretion events with $\delta\geq0.08$
had a very similar cosmological evolution to the $\delta\geq0.08$ curves in Fig.\ref{dm_by_m}.
The minor merger and diffuse accretion events with $\delta\leq0.02$
also showed a similar behaviour to the $\delta\leq0.02$ curves in Fig.\ref{dm_by_m}, except 
there were more minor mergers at higher 
redshift for all mass curves.
These features do not affect our conclusions in Section \ref{Croton_section}, however.

Fig.\ref{mass_func} shows the shift from major-merger and diffuse- 
dominated growth at high redshift to minor-merger and diffuse- dominated growth at low redshift, 
more clearly. The linestyles have the same meaning as in Fig.\ref{dm_by_m}, 
except we also include the halos with $0.02<\delta<0.08$, shown by the dotted lines. 
It can be seen that minor-mergers and diffuse accretion events
start to significantly contribute to growth for 
$z<0.5$, and by $z=0$ drive accretion onto all halo masses.

Qualitatively we find very similar results to Figs. \ref{dm_by_m} and \ref{mass_func} when 
halos are binned in $\mu$ ($\equiv\dot{M}/M$) instead of $\delta$, but the thin major-merger curves 
in Figs. \ref{dm_by_m} and \ref{mass_func} decouple from the thick curves at later epochs, for all masses. 
This is probably because in transitioning from $\delta$ to $\mu$, one must divide $\delta$ 
by the time interval during which mass is accreted, and at higher redshifts this time 
interval is smaller (time is not a linear function of redshift) and $\mu$ is hence 
larger than it is for a given $\delta$ onto a halo of fixed mass at lower redshift. 

\begin{figure}
\centering
\includegraphics[width=0.95\columnwidth,height=7cm]{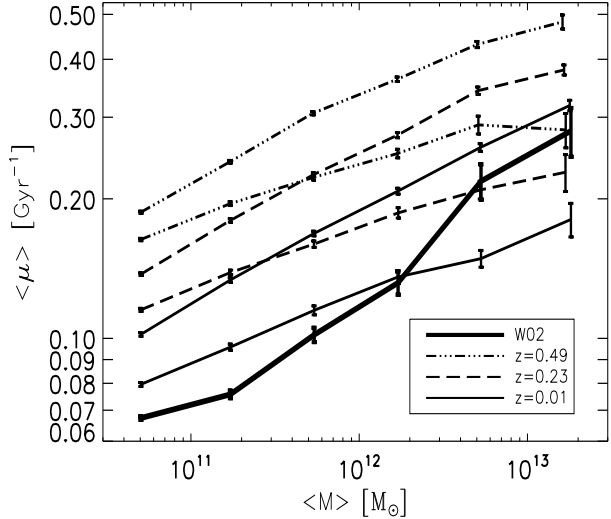}
\caption{The mean specific accretion rate onto halos and subhalos using the halosub method, 
plotted as a function of object mass for three redshifts corresponding to $z=0.49$ 
(triple-dot-dashed lines), $z=0.23$ (dashed lines) and $z=0.01$ (solid lines). 
Lines of a given linestyle from bottom to top represent the accretion onto the 
AHOP halos and the MSM halos and subhalos respectively. The thick line shows the W02 result, 
obtained by using equation (\ref{eWech}) and equation (\ref{eAlpha}) at $z=0.01$.}
\label{msm-hop-plots}
\end{figure}

\begin{figure}
\centering
\includegraphics[width=0.95\columnwidth,height=7cm]{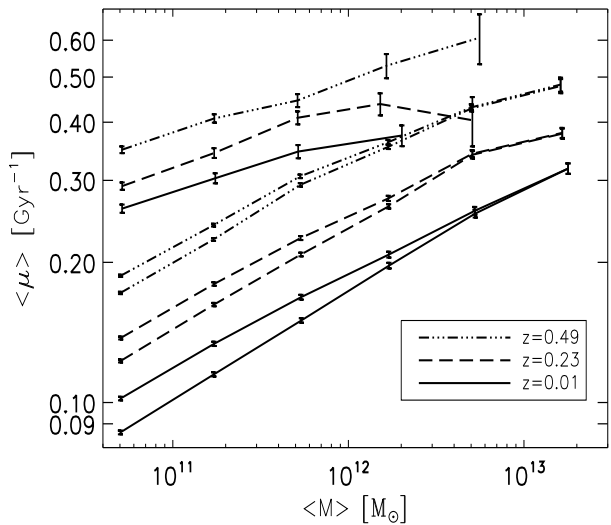}
\caption{The mean specific accretion rate as a function of mass shown for $z=0.49$ 
(triple-dot-dashed lines), $z=0.23$ (dashed lines) and $z=0.01$ (solid lines) 
using the halosub accretion method. For a given linestyle, the bottom line shows 
the MSM halos, the middle line shows the MSM halos and subhalos and the top line shows the MSM subhalos.}  
\label{accn-components}
\end{figure}

\subsection{Accretion onto subhalos}

In this section, the AHOP halos are resolved into constituent MSM halos and subhalos and the 
halosub method is applied to these resolved structures to account for accretion onto 
objects in groups and clusters. We begin by comparing the AHOP halo and MSM halo and 
subhalo specific accretion rates with the results found in the W02 simulation study. 
The mass of a halo or subhalo is henceforth denoted by $M$, in contrast with the previous 
section which only recorded accretion onto halos with mass $M_{H}$.

\subsubsection{Comparing the halosub accretion algorithm with W02}

Fig.\ref{msm-hop-plots} plots the average specific accretion rate for all bound objects 
from the simulation as a function of average object mass for redshifts 
corresponding to $z=0.49$ (triple-dot-dashed lines), $z=0.23$ (dashed lines) 
and $z=0.01$ (solid lines). These redshifts have been chosen because the 
epoch of cluster formation is $z<1$. The lines of a given linestyle from 
bottom to top represent the accretion onto the AHOP halos and MSM halos 
and subhalos respectively. The thick line shows the W02 result at $z=0.01$ 
using equation (\ref{eWech}) (strictly, equation (\ref{eWech}) holds at $z=0$ 
but we cannot use our anomalous detection method at this redshift). 
The W02 result was calculated by binning in mass each $z=0.01$ bound AHOP halo 
accretor and computing the corresponding average W02 $\alpha$ parameter 
in equation (\ref{eAlpha}) for each mass bin ($\alpha$ is inversely 
proportional to halo formation redshift). 

The specific accretion rate onto the MSM objects is systematically larger than the AHOP specific accretion
rates at every mass when considering a given redshift. 
The MSM method resolves the substructure that has been averaged out in the AHOP halo, 
so the main MSM host halo and subhalos are individually less massive than the AHOP counterpart. 
The offset with MSM is probably caused by dividing by the larger AHOP mass, and 
this offset increases with increasing mass because at larger masses subhalos occupy a larger 
fraction of the total AHOP mass. 
The mass difference between AHOP and the main host MSM halo therefore increases with 
increasing AHOP mass (and there are more detected halos than subhalos at a given redshift
in the simulation, so the halos dominate the MSM halo and subhalo accretion signal). 

W02 fitted the accretion trajectories of their $z=0$ halos averaged over environment in a 
WMAP1 cosmology and so their result can be directly tested against the AHOP curve 
at $z=0.01$ which also averaged over environment, but in a universe with a WMAP3 
cosmology (W02 argue that their fitting formula does not depend on the chosen cosmology). 
We find that the W02 specific accretion rate has a stronger mass dependence than found 
for the AHOP halos in this study and so for the large galaxy- and group- sized dark halos, 
overpredicts the specific accretion rate by a factor of $\sim 1.5$.

Recent studies have shown that some halo-finding algorithms can lead to large
uncertainties in the halo accretion rate \citep{Genel09,Hopkins10}. 
The disagreement across mass with W02 in Fig.\ref{msm-hop-plots}, however, does not result from 
differences in halo-finder: the AHOP algorithm is 
very similar to the modified bound density maxima technique of \cite{Bullock01} used in W02.
The disagreement most likely arises because W02 impose different criteria
to identify the main son and main father.
They adopt a policy, in some cases, of tracking the single most bound particle, 
which is misleading as the trajectory essentially performs a random walk across time. 
By constrast, we rigorously identify false merger candidates and adopt an accretion algorithm that
tracks channels which donate/receive the most mass (and recall that by allowing a 
bound object below the mass threshold to be a main father, we do not bias the accretion events). 
Our method hence avoids using ad-hoc criteria.

\begin{figure*}
\includegraphics{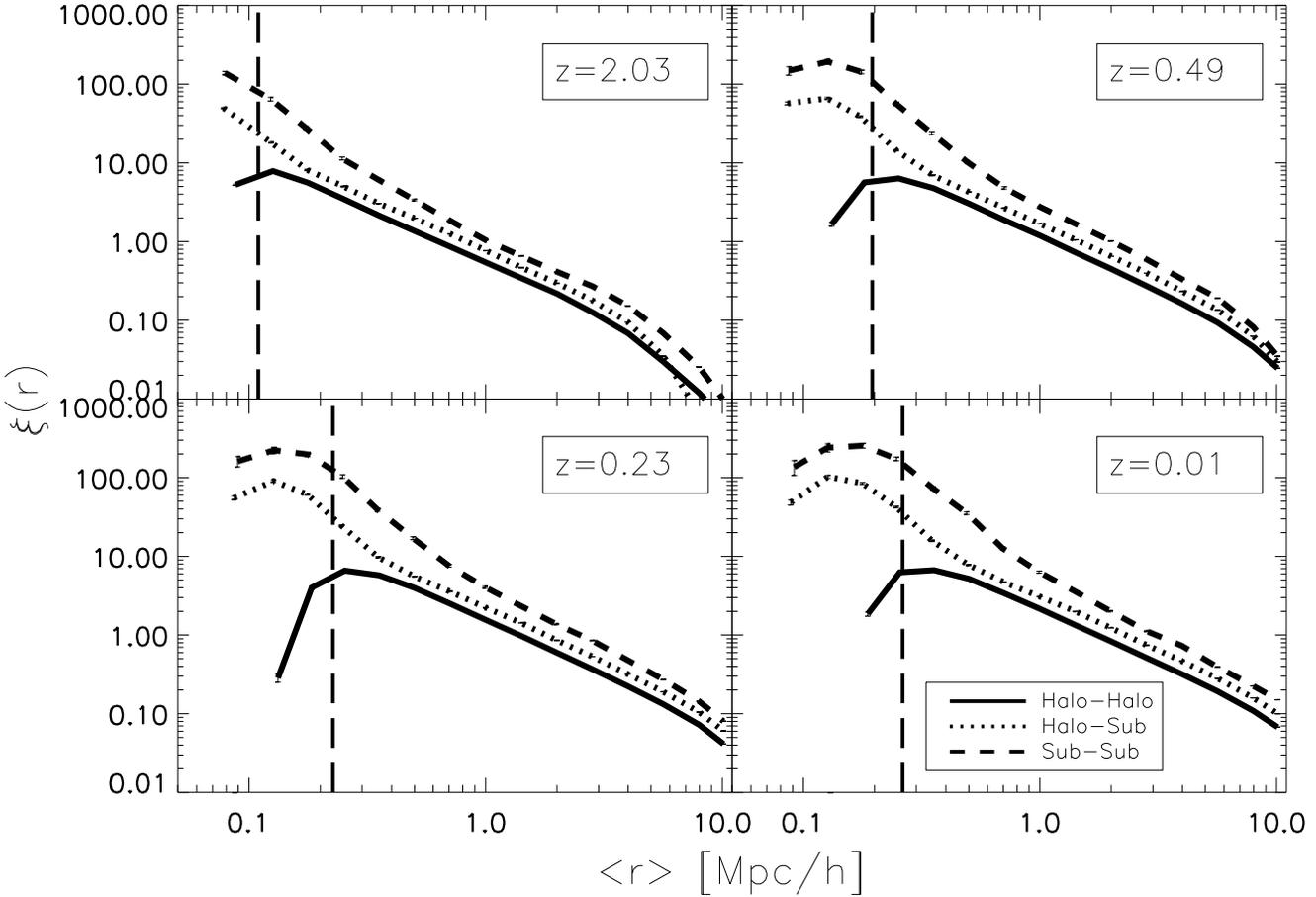}
\caption{The two-point correlation function plotted as a function of the 
mean inter-object separation (in physical coordinates) for four redshifts 
corresponding to $z=2.03, 0.49, 0.23$ and $z=0.01$. The lines in each panel 
represent the bound MSM halo-halo (solid), halo-subhalo (dotted) 
and subhalo-subhalo (short-dashed) accretors. 
The vertical long-dashed lines represent an estimate of the resolution limit in $r$ at each redshift.}
\label{xi_plots}
\end{figure*}

\begin{figure*}
\includegraphics{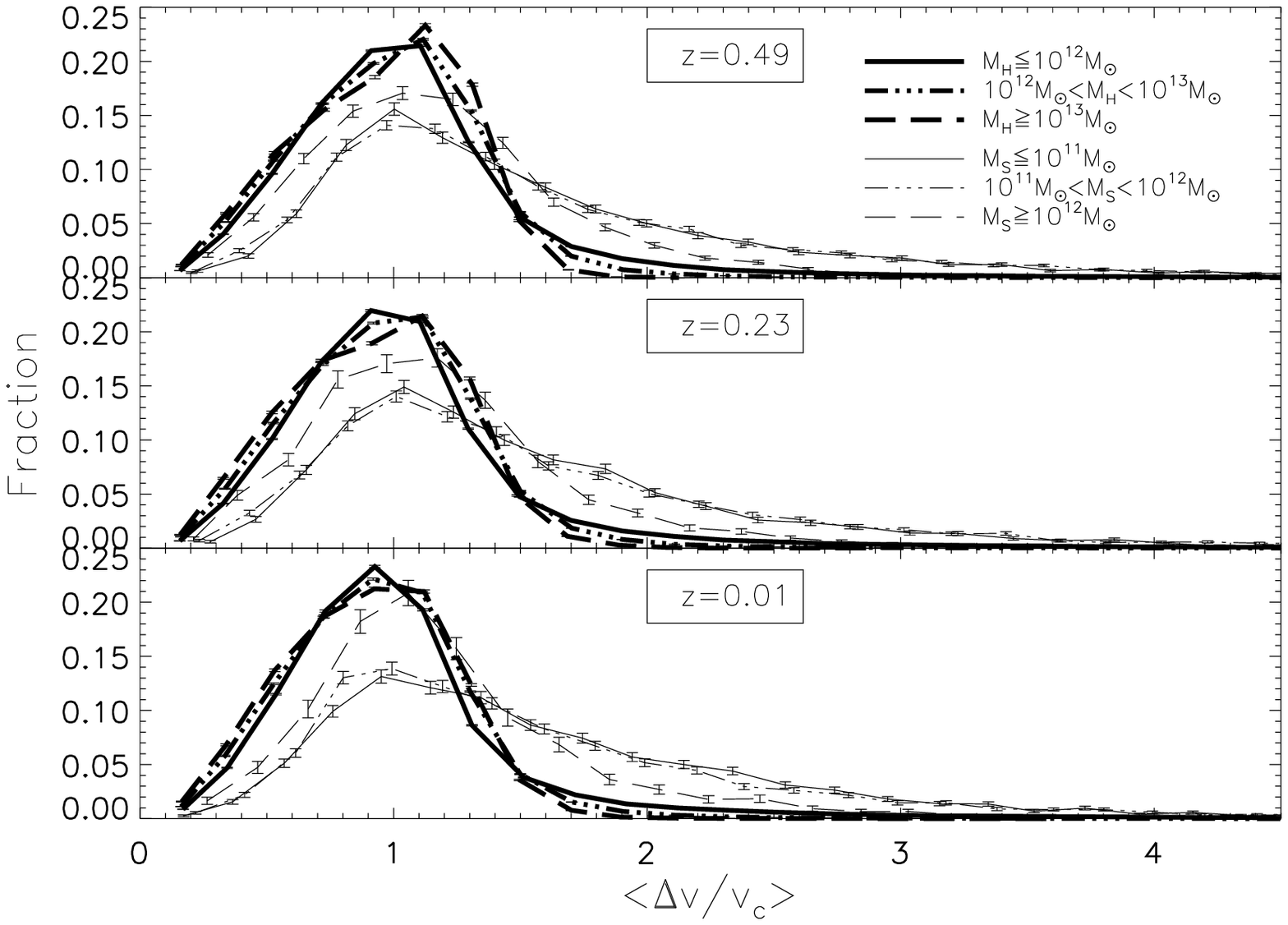}
\caption{Distributions of $\Delta v/v_{c}$ for the halo (thick lines) 
and subhalo (thin lines) accretors, where $\Delta v$ represents 
the relative velocity between an accretor's main father
and one of its other progenitors, and $v_{c}$ is the accretor's circular
velocity. The halo and subhalo accretors are divided into
different mass bins, shown by the ranges of $M_{H}$ and $M_{S}$ respectively,
and correspond to the same objects shown in Fig.\ref{accn-components}.}
\label{vrel_fig}
\end{figure*}

\subsubsection{Accretion onto MSM halos and subhalos}

Fig.\ref{accn-components} shows the specific accretion rate from bottom to top of MSM halos, 
MSM halos and subhalos, and MSM subhalos with the linestyles having the same meaning as in 
Fig.\ref{msm-hop-plots}. The average specific accretion rates onto halos ($\mu_{H}$) and subhalos ($\mu_{S}$) have 
weak mass dependencies for each of the redshifts shown: 
$\langle\mu_{H}\rangle\propto M^{0.2}$ and $\langle\mu_{S}\rangle\propto M^{0.1}$ at $z=0.01$, for example. 
Each of the halo, halo and subhalo, 
and subhalo curves shift downwards with decreasing redshift:
the average specific accretion rate onto a subhalo at $z=0.49$ is a factor of $1.3-1.4$ 
greater than at $z=0.01$, for example. 
Major merger and diffuse accretion events at higher redshifts, 
when the universe was more dense, are more prominent.

Fig.\ref{accn-components} also reveals that the subhalo accretors
(and this includes the subhalos with a zero accretion rate) accrete at a larger rate, on average, 
than the halo accretors for $z<0.5$ at the mass scales shown. 
This, however, only causes a modest shift 
from the halo curve to the halo and subhalo curve at each 
redshift, because there are more halo accretors than 
subhalo accretors in the simulation, 
indicating that the subhalos are not responsible for the 
AHOP to MSM shift in accretion at each redshift in Fig.\ref{msm-hop-plots}.
The enhanced accretion onto subhalos can be understood by examining
their mutual clustering and the relative velocity of their progenitors compared
to their internal velocity, and both of these processes 
are discussed in the following sections.

\subsubsection{The clustering of halos and subhalos}\label{cluster_section}

The main aim of this study is to investigate whether there is a 
relationship between the rate at which objects accrete mass 
and their environment and so in this section the clustering 
properties of halos and subhalos at different redshifts are 
examined. In the following section we specifically target 
accretors in different cluster-scale environments.
 
Fig.\ref{xi_plots} shows the two-point correlation function, $\xi$, for 
the MSM accretors from the simulation as a function of the physical separation 
distance $r$, at the same three redshifts shown in Figs. \ref{msm-hop-plots} 
and \ref{accn-components} and at a much higher redshift of $\sim2$. 
The \cite{Landy} $\hat{w}_{4}$ estimator was used to compute $\xi$, requiring 
random catalogues for each redshift. Our catalogues sampled $300000$ 
objects at each redshift and were hence larger than the corresponding total number 
of detected halos and subhalos ($z=2.03:156120; z=0.49:211537; z=0.23:216232; z=0:223781$). 
For each panel in Fig.\ref{xi_plots}, the solid lines represent the halo-halo pairs, the dotted lines 
represent the halo-subhalo pairs and the dashed lines represent the subhalo-subhalo pairs.
Only the clustering of bound accretors was measured:
halo-subhalo pairs correspond to the clustering of all bound halo accretors with all 
bound subhalo accretors, for example. 
The vertical dashed lines show the average total diameter of an object at the redshift in question and 
represent an estimate of the resolution limit in $r$. 

Fig.\ref{xi_plots} demonstrates that subhalo-subhalo pairings are a factor 
of $\sim2$ more clustered than halo-halo pairings at large physical scales
at low redshift. 
This factor increases to $\sim10-15$ at lower separation scales:
subhalos, by definition, reside within halos and so cluster
more strongly at small scales. The drop-off in 
clustering amplitude at the lowest scales should be ignored as this occurs at scales 
that are below the estimated resolution limit.

The subhalo-subhalo correlation function is the sum of two terms: the first describes the clustering of subhalos
within the same host and the second describes the clustering of subhalos that belong 
to different hosts. For small separations, the subhalo-subhalo correlation function 
has a strong contribution from pairs of subhalos in the same host. 
The clustering of halo-halo pairings is lower at these scales 
because these scales approach the size of halos, and so 
it is less common to find two halos close to each other without one or both member(s) 
of the pair being a subhalo.
At larger scales, subhalos belonging
to different hosts contribute strongly to the subhalo-subhalo clustering strength. 

The clustering amplitudes of the three curves also evolve with redshift: 
the correlation length of the subhalo-subhalo curve increases by a 
factor $\sim3$ towards $z=0$, for example. This is probably because at lower redshift there are more dense 
clusters and more subhalos within a given host in the simulation, hence there is a stronger contribution 
to the subhalo-subhalo clustering amplitude than at higher redshift at the separation scales shown. 

\subsubsection{Measuring the relative velocities between the accretors' progenitors}

Having established that subhalos at sub-cluster scales are more clustered than
halos, especially at small scales, we now examine the distributions of
$\Delta v/v_{c}$, where $\Delta v$ represents 
the relative velocity between an accretor's 
main father and one of its other progenitors, and $v_{c}$ is the accretor's circular velocity. 
If $\Delta v/v_{c}$ tends to be smaller, on average, for subhalo accretors
than halo accretors for example, 
then accretion onto halos will tend to be more suppressed than
accretion onto subhalos. 
Fig.\ref{vrel_fig} shows the distributions of this ratio for halos (thick lines)
and subhalos (thin lines) at the same redshifts shown in Fig.\ref{accn-components}.
The $\Delta v/v_{c}$ ratio was computed for each progenitor $k$
(not equal to the main father $j$) of a given accretor:
each particle accreted from the background was counted 
as an individual relative velocity event, as was each halo/subhalo
progenitor. So if, for example, an accretor has a main father $j$, a father
$k$, and also accretes two particles from the background, $m$ and $n$, then
three separate relative velocities with respect to $j$ are computed for that accretor. The accretors
were binned in mass, and the different halo and subhalo mass bins are
shown by the ranges of $M_{H}$ and $M_{S}$ in Fig.\ref{vrel_fig}, respectively.

It can be seen from Fig.\ref{vrel_fig} that 
the distributions of $\Delta v/v_{c}$ 
for the halo and subhalo accretors are similar: they
depend quite weakly on mass and their peaks coincide.

\subsubsection{Revisiting the enhanced accretion onto subhalos in Fig.\ref{accn-components}}

It is well established that in simulations, after infall, subhalos 
experience mass loss via tidal stripping, tidal heating and
disk shocking \citep{Gnedin,Dekel03,Taylor_Babul,Onghia}, and have a large velocity dispersion 
that scales with their host's mass. 
Mass stripping from an object in this dark-matter-only study is recorded as zero accretion, and 
so one would perhaps expect subhalos to be accreting at low rates, on average.
We have found, however, that most of the subhalo accretors in
the simulation at $z<0.5$ reside in the outer regions of their host,
with $\sim70\%$ located beyond their host's virial radius. 
(The halo virial radius roughly corresponds to $r_{200}$, which encloses the region
within which the halo density is at least $200$ times the critical density of the universe.)
Most of these subhalos have therefore probably not been significantly stripped of their mass.
Infact, we find the opposite trend in Fig.\ref{accn-components}:
subhalos of a given mass in the simulation have a larger rate of accretion,
on average, than halos of the same mass. 
Having demonstrated that there is no significant difference between the 
halo and subhalo accretor distributions of $\Delta v/v_{c}$,
we conclude that the enhanced subhalo accretion rates
are driven by the very frequent interactions between
subhalos of the same host at small scales (Fig.\ref{xi_plots}). 
Halos are less clustered at small scales
and so accrete at lower rates, on average.

\begin{figure*}
\includegraphics{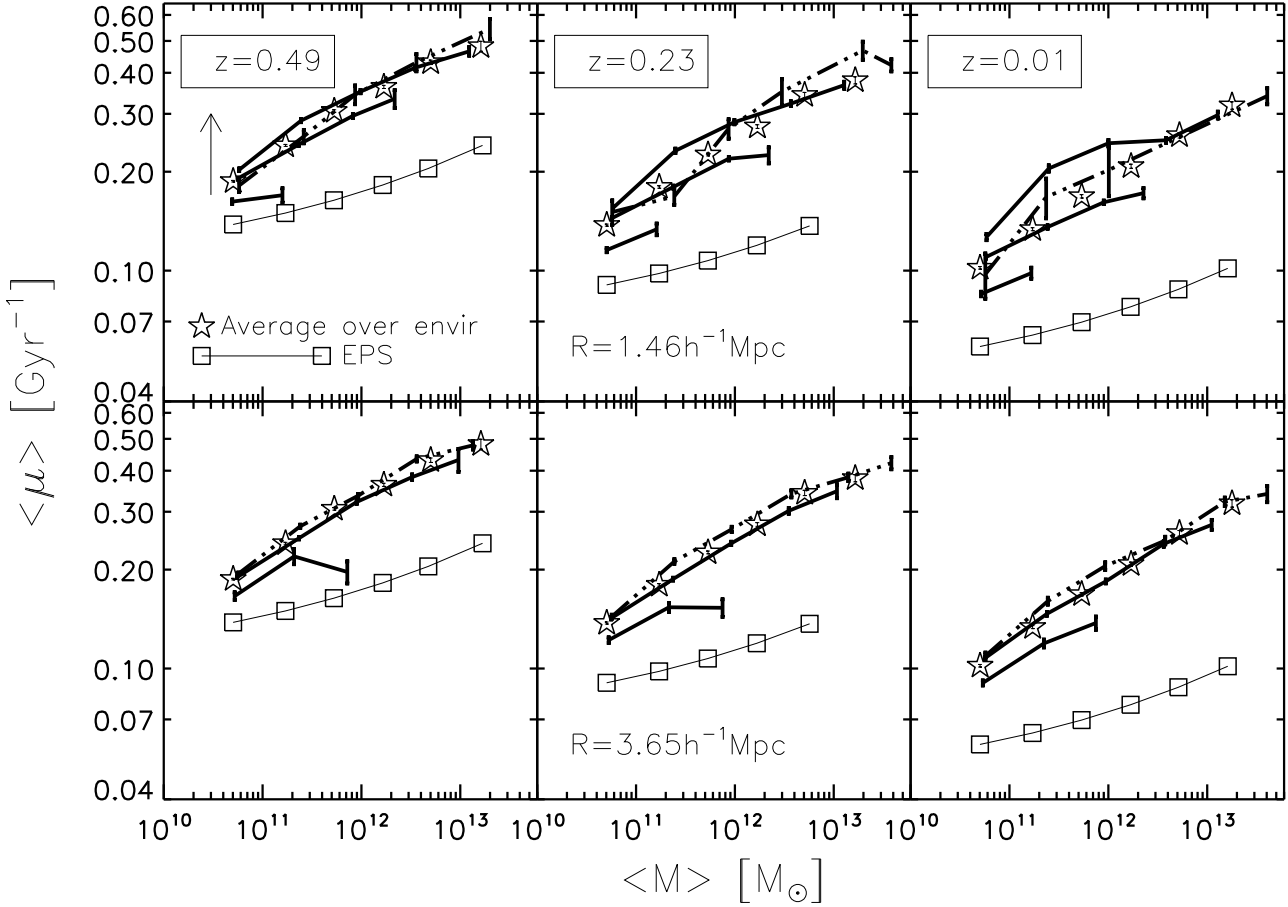}
\caption{$\langle \mu \rangle$ plotted as a function of object mass ($M$) and environment mass 
($M_{E}$) for MSM-detected objects. The lines in row one, from bottom to top, 
represent $M_{E}<10^{11.5}M_{\odot}$, $10^{11.5}M_{\odot}\leq M_{E}<10^{12.5}M_{\odot}$, 
$10^{12.5}M_{\odot}\leq M_{E}<10^{13.5}M_{\odot}$ (solid lines) and $10^{13.5}M_{\odot}\leq M_{E}<10^{14.5}M_{\odot}$ 
(triple-dot-dashed lines) using a sphere radius ($R$) of $1.46h^{-1}$Mpc. The second row shows the results 
for a sphere radius of $3.65h^{-1}$Mpc for environment mass bins $M_{E}<10^{12.5}M_{\odot}$, 
$10^{12.5}M_{\odot}\leq M_{E}< 10^{13.5}M_{\odot}$ (solid lines)
and $10^{13.5}\leq M_{E}<10^{14.5}M_{\odot}$ (triple-dot-dashed lines). 
The vertical arrow indicates the direction of increasing environment mass for both rows, 
with the exception of the highest environment mass bins in row one, which mostly lie 
beneath the second highest environment bins.
The open squares joined by solid lines illustrate the EPS result using equation (\ref{eMill}) and the open stars 
show accretion onto the MSM halos and subhalos, independent of their environment. 
Columns one, two and three correspond to $z=0.49, z=0.23$ and $z=0.01$.}
\label{environment_plots}
\end{figure*}

\begin{figure*}
\includegraphics{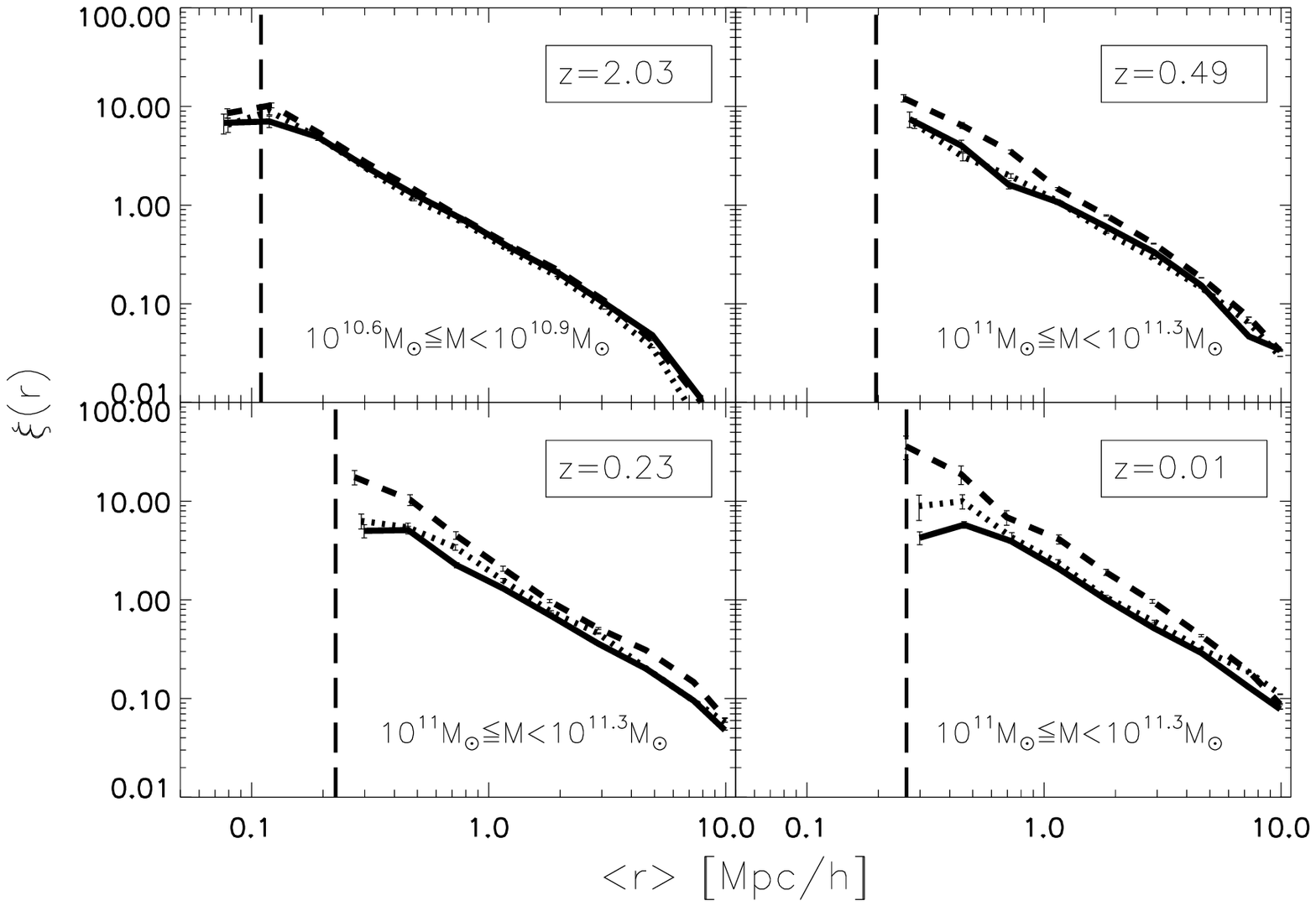}
\caption{The two-point correlation function plotted as a function of the mean 
inter-object separation (in physical coordinates) for the four redshifts 
corresponding to $z=2.03, 0.49, 0.23$ and $z=0.01$. The $z=2.03$ panel 
shows the MSM objects whose masses satisfy $10^{10.6}M_{\odot}\leq M<10^{10.9}M_{\odot}$ 
with $\mu<0.35$Gyr$^{-1}$ (solid), $0.35$Gyr$^{-1}\leq\mu<0.6$Gyr$^{-1}$ (dotted) 
and $\mu\geq0.6$Gyr$^{-1}$ (short-dashed). The lower redshift panels show the MSM objects 
whose masses satisfy $10^{11}M_{\odot}\leq M<10^{11.3}M_{\odot}$ with $\mu<0.1$Gyr$^{-1}$ 
(solid), $0.1$Gyr$^{-1}\leq\mu<0.2$Gyr$^{-1}$ (dotted) and $\mu\geq0.2$Gyr$^{-1}$ 
(short-dashed). The vertical long-dashed lines represent an estimate of the 
resolution limit in the separation scale at each redshift.}
\label{xi_percival_plot}
\end{figure*}

\subsection{Halo and Subhalo environment}\label{envir_section}

In this section we specifically target the effect an object's environment at cluster scales has 
on the rate at which it accretes mass. There are two popular, independent measures 
of environment in the literature; the overdensity $\delta_{R}(\textbf{x})$ in a sphere of
radius $R$ \citep{Lemson, Wang} 
and halo bias \citep{Sheth04, Gao07}. We adopt two similar measures of an object's environment:
the first defines an environment mass within a cluster-sized sphere
and the second uses the two-point correlation function.

\subsubsection{Environment mass}\label{envirmass}

We have defined the environment of a halo and a subhalo as the total mass, $M_{E}$, 
contained within a sphere of radius $R$ centred on the object of interest. 
$M_{E}$ includes the mass of all those objects whose centres lie within the sphere 
as well as the mass of the object the sphere is centred on. 
We consider spheres of radii $R=1.46h^{-1}$Mpc and $R=3.65h^{-1}$Mpc because 
a) these scales represent both typical clusters and much larger clusters and 
b) various authors have 
found that the dependence of some halo properties 
on environment, such as halo formation redshift, are sensitive to the 
choice of sphere radius \citep{Lemson, Harker, Hahn}.
Both of these environment mass definitions are applied to 
each bound accretor at the redshift under consideration, with 
only bound accretors having a recorded $M_{E}$ value. Unbound objects and 
resolved objects with $M\leq40M_{p}$ 
are not, however, excluded from the sample as these 
objects could be part of a bound object's environment.

The first row of Fig.\ref{environment_plots} plots the specific 
accretion rate onto halos and subhalos as a function of average 
object mass ($M$) and average environment mass ($M_{E}$) for $z=0.49$ 
(first column), $z=0.23$ (second column) and $z=0.01$ (third column) 
using a sphere radius of $1.46h^{-1}$Mpc. The 
second row of Fig.\ref{environment_plots} shows the results using a larger sphere 
radius of $3.65h^{-1}$Mpc at the same three redshifts. 

The solid lines represent the environment mass bins which from bottom to 
top for the first row are: $M_{E}<10^{11.5}M_{\odot}$, 
$10^{11.5}M_{\odot}\leq M_{E}<10^{12.5}M_{\odot}$ and $10^{12.5}M_{\odot}\leq M_{E}< 10^{13.5}M_{\odot}$. 
The triple-dot-dashed line shows the largest environment mass bin of $10^{13.5}M_{\odot}\leq M_{E}<10^{14.5}M_{\odot}$. 
For the larger scale environments in the second row (from bottom to top): $M_{E}<10^{12.5}M_{\odot}$, 
$10^{12.5}M_{\odot}\leq M_{E}<10^{13.5}M_{\odot}$ (solid lines) and $10^{13.5}M_{\odot}\leq M_{E}<10^{14.5}M_{\odot}$ 
(triple-dot-dashed lines). The vertical arrow shows the direction of increasing environment for all panels, 
with the exception of the largest environment mass bins in the first row, which mostly lie beneath the 
second largest environment bins. The stars in each panel represent the accretion onto MSM halos 
and subhalos independent of their environment and the squares joined 
by solid lines show the EPS results. 

The relationships found in the previous sections are preserved in Fig.\ref{environment_plots}: 
the specific accretion rate increases with object mass for objects in most environments 
and decreases towards $z=0$ (as was shown in Fig.\ref{accn-components}), and EPS consistently underestimates 
the mass accreted onto all object masses (as was shown for halos at $z<1$ in Fig.\ref{halo_accn}). 
The most striking feature of Fig.\ref{environment_plots}, however, is that objects of a 
given mass residing in more massive environments do not accrete
at a particularly enhanced rate compared with objects of the same mass 
in much lower mass environments. This suggests that the specific 
accretion rate onto halos and subhalos does not depend strongly on environment at cluster scales. 
Objects in cluster mass environments shown in the first row (triple-dot-dashed lines)
mostly accrete less mass than in lower mass 
environments, but the number of objects in cluster mass surroundings is limited by the choice of sphere radius. 
This effect is not seen for the larger-scale 
environments shown in the second row, for example, where merging between subhalos on the outskirts of the host halo 
is probably driving accretion (but only at a slightly higher overall rate). The second row 
shows that the specific accretion rate only depends weakly on environment
at larger scales that probe the outermost regions of clusters.
This weak environment dependence in rows $1$ and $2$ therefore seems to suggest that
the increased interaction rates of halos in group- and cluster- mass environments
are not sufficiently large enough to significantly overcome the large halo relative
velocities, resulting in only a modest net increase in accretion.

Halos dominate the accretion signals in Fig.\ref{environment_plots}, but we find 
the same trends at each of the chosen redshifts when just subhalos are plotted as a function of their
mass and environment mass. There are two differences, however: 
the subhalos a) accrete at higher rates and 
b) reside only in larger mass environments. The subhalo curves 
have been omitted in Fig.\ref{environment_plots} for clarity.

Other authors have quantified environment by computing the overdensity $\delta_{R}$ in 
a sphere of radius $R$, rather than the mass \citep{Lemson,Harker,Hahn,Fakhouri,Fakhouri_diffuse}. 
We therefore calculated a weighted environment density for each halo and subhalo accretor
by using the standard SPH cubic spline window function \citep{Monaghan} 
which weights the mass contributions from objects
close to the centre of the sphere more strongly than those further away. 
\cite{Fakhouri} showed that for halos more massive than $10^{14}M_{\odot}$, the density of
the object the sphere is centred on starts to dominate the contributions to $\delta_{R}$,
and so the central object's contribution was therefore both included
and excluded in two separate weighted environment 
density measures.
When binned in environment density, the same weak environment dependence 
as in Fig.\ref{environment_plots} was found in both cases.

\subsubsection{Clustering in different accretion schemes}

In this section we use the correlation function as an alternative means 
to Section \ref{envirmass} of measuring an object's environment, except we
do not restrict our analysis to just cluster scales of a few Mpc. 
We consider samples of objects with very similar masses
at different redshifts and examine whether objects 
of a given mass which accrete at larger rates have a 
larger clustering amplitude. This also tests the work by \cite{Percival}, 
who found that at $z=2$ halos of a given mass accreting at different 
rates do not cluster differently.

The $z=2.03$ panel in Fig.\ref{xi_percival_plot} shows the correlation 
function for all those objects whose mass satisfies 
$10^{10.6}M_{\odot}\leq M<10^{10.9}M_{\odot}$ with $\mu<0.35$Gyr$^{-1}$ (solid), 
$0.35$Gyr$^{-1}\leq\mu<0.6$Gyr$^{-1}$ (dotted) and $\mu\geq0.6$Gyr$^{-1}$ (dashed). 
The lower redshift panels show the correlation function for objects whose mass 
satisfies $10^{11}M_{\odot}\leq M< 10^{11.3}M_{\odot}$ with $\mu<0.1$Gyr$^{-1}$ 
(solid), $0.1$Gyr$^{-1}\leq\mu<0.2$Gyr$^{-1}$ (dotted) and $\mu\geq0.2$Gyr$^{-1}$ 
(dashed). The mass interval for $z<0.5$ has been chosen because
it lies below the break mass, $M_{\star}$, in the mass function 
at these redshifts and so we do not bias $\mu$. For comparison, the mass interval
in the $z\sim2$ panel lies closer to $M_{\star}$.
The vertical dashed lines represent an estimate of the resolution limit in the separation 
scale (same as the vertical dashed lines in Fig.\ref{xi_plots}).

At well resolved non-linear small scales for $z<0.5$, objects with high specific accretion rates 
are up to a factor of $\sim3$ more clustered than the lower accreting objects,
whereas at larger linear scales the difference in clustering between different accretors is much smaller. 
For the cluster-scale environments of the first row of Fig.\ref{environment_plots}, corresponding to 
an $r$ value of $2.92h^{-1}$Mpc, there is a weak environment dependence, with objects of larger 
$\mu$ being slightly more clustered. Fig.\ref{xi_percival_plot} therefore provides further 
evidence that the mass accreted onto halos and subhalos of a given mass weakly 
depends on their environment at cluster scales. 

In contrast to the $z<0.5$ behaviour, there is very little 
difference in clustering between different accretors with 
$10^{10.6}M_{\odot}\leq M<10^{10.9}M_{\odot}$ at $z\sim2$ and 
this holds for both the linear and non-linear scales shown. 
We therefore agree with the conclusions of \cite{Percival} at 
$z\sim2$ but show that they break down at $z<0.5$, where there 
is a larger difference in clustering between high accretors and low accretors 
of a given mass at all scales.

\section{Discussion}\label{discussion_section}

\subsection{Disagreement with EPS theory}\label{EPS_section}

Despite its success at reproducing the dark halo mass function in simulations, we find that the analytic
EPS calculation shows significant departures from the halo accretion rates found in our simulation at both low
and high redshift (Fig.\ref{halo_accn}). 
This simulation study, however, is not the first to report disagreement with EPS theory at high redshift: 
\cite{Cohn08} examined the accretion onto halos of mass $M_{H}=5-8\times10^{8}h^{-1}$M$_{\odot}$ at $z=10$ and found
 that EPS overestimated the halo accretion rate by a factor $\sim 1.5$ (using a lookback time of $50$ Myrs). 
Fig.\ref{halo_accn} shows a similar behaviour, with EPS overpredicting the accretion rate onto halos of mass 
$M_{H}\sim10^{10.7}M_{\odot}$ by a factor of $\sim2$ at $z=8$. One might expect EPS to overestimate accretion 
onto halos at high redshift because it assumes that collapse is spherical and that the density barrier is fixed 
in height \citep{Lacey} whereas it has been shown that allowing for ellipsoidal collapse and treating the critical 
density contrast for collapse as a free parameter better reproduces the N-body halo mass function \citep{Sheth01,Sheth02}. 
This modification reduces the critical density contrast for collapse by a factor of $\sqrt{0.7}$ (M06) which reduces 
$f(M_{H})$ in equation (\ref{eMill}) by the same factor, causing a slight shift in the dashed curves in 
Fig.\ref{halo_accn} but otherwise having no effect on the redshift or mass dependence. 

The disagreement might arise because EPS theory is only approximate: (i) it assumes spherical collapse, 
whereas halos in dark-matter-only simulations are triaxial; (ii) it contains no dynamical information,
and so is unable, for example, to account for mass 
being stripped from one halo and then being accreted onto another; 
(iii) it cannot account for accretion onto substructures; and (iv) it averages over halo environment. 
The latter restrictions are particularly problematic in the non-linear regime at $z<1$, when 
accretion onto structures embedded within clusters is of interest (Fig.\ref{environment_plots}). 
Recent attempts to incorporate an environment dependence into the EPS excursion set theory 
\citep{Maultbetsch,Sandvik,Zentner,Desjacques} could modify equation (\ref{eMill}) which might 
result in better agreement with our simulation results for $z<1$ in Fig.\ref{environment_plots}. 
\cite{Benson05} highlighted further weaknesses with the EPS formalism that could also account for the offset 
in Figs. \ref{halo_accn} and \ref{environment_plots}. 
They showed that the \cite{Lacey} EPS formula yields merger rates that 
are not symmetric under exchange of halo masses, and which do not predict the correct evolution of 
the Press-Schechter mass distribution, indicating that constructed EPS merger trees are fundamentally flawed.

Despite these limitations, the gradients of the EPS curves in Fig.\ref{halo_accn} are 
only slightly steeper than the corresponding simulation
curves. This implies that the \cite{Lacey} EPS formalism may only require minor adjustment
to agree more closely with the simulation trajectories across mass and redshift.

\subsection{The weak relationship between accretion rate and environment at cluster scales}

By quantifying accretion onto substructures embedded in groups and clusters, we have moved beyond the limited 
predictive power of the EPS formalism. Fig.\ref{accn-components} demonstrates that subhalos accrete 
at larger rates than halos of the same mass, on average, in the simulation
(by a factor of $\sim3$ for the lowest mass subhalos at $z=0$). 
At first glance this appears to contradict recent
claims: \cite{Angulo} and \cite{Hester}, for example, have
shown that subhalo-subhalo mergers
are rare and that subhalos are severely stripped of mass, 
which probably means that the accretion 
rates onto their subhalos are likely to be low.
The subhalo accretors at low redshift in this study, however, form a subsample of
subhalos that are safely above the mass resolution limit and that are mostly located
at large distances from their host's centre, 
with $\sim70\%$ residing beyond their host's virial radius.
(The halo virial radius approximately encloses
the region within which the halo density is at least $200$ times
the critical density of the universe).
These subhalos are probably not therefore being significantly stripped of mass,
unlike the subhalos in recent studies. The mass-selected
nature of our subhalo accretors and the different spatial 
distribution within their host 
are therefore the most likely causes 
of the apparent accretion rate discrepancy with the studies mentioned above. 
We have further shown that the subhalo accretors in this study are more clustered
than the halo accretors at small scales (Fig.\ref{xi_plots}) 
and that there is no significant difference between the
distributions of $\Delta v/v_{c}$, where $\Delta v$ is the relative velocity 
between an accretor's main father and one of its other progenitors, and
$v_{c}$ is the accretor's circular velocity.
The high subhalo accretion rates are therefore likely to be 
driven by the very frequent interactions at small scales
with other subhalos of the same host. 

One might expect the accretion rate onto halos and subhalos to depend strongly on environment at 
larger, cluster-sized scales given the increased
rate of interactions in dense environments, but only a weak dependence 
is found (Figs. \ref{environment_plots} and \ref{xi_percival_plot}). 
The subhalo accretors reside in
only the most massive environments and probably accrete mostly locally 
from their nearby subhalo neighbours rather than their host, 
and so this is a possible explanation for their
weak relationship between accretion rate and environment.
One likely explanation for halos is that
the increased interaction rates of halos in group- and cluster- mass environments
are not sufficiently large enough to significantly overcome the large halo relative
velocities, resulting in only a modest net increase in accretion at cluster scales.

\cite{Fakhouri_diffuse} examined the environment dependence of
accretion onto high mass halos ($M_{H}>10^{12}M_{\odot}$)
from the Millennium simulation
and found a weak, negative correlation for galaxy-mass halos. 
We find a weak but positive dependence for all object masses in Fig.\ref{environment_plots}.
Our analysis, which extends theirs by accounting for accretion onto substructures,
has a different expression for the mass accretion rate but 
we have found little difference in the results obtained from using the two expressions
for bound objects. 
The obvious source of the discrepancy is therefore 
the method used to identify anomalies. 
\cite{Genel09} have highlighted some fundamental problems 
with the `stitching' algorithm used by \cite{Fakhouri_diffuse} to remove anomalous events,
demonstrating that it can lead to a double counting of mergers
and to a false counting of anomalous events as mergers. 
They show that this overestimation of the merger rate 
is particularly problematic for minor-mergers.
Predicting the effects that overestimating the merger-rate 
has on the accretion rate and how this varies as a function of environment is not trivial,
but differences between the anomalous event detection methods could explain
the difference in the sign of the trend between accretion rate and environment.

The $z=2.03$ panel in Fig.\ref{xi_percival_plot} reveals that at higher redshift
when halos far outnumber subhalos in the simulation, 
the rate of accretion onto halos is independent of environment, confirming the \cite{Percival}
result. The \cite{Percival} study examined the difference 
in clustering at $z=2$ between halos of a given mass accreting at different rates. 
They considered several mass intervals 
ranging from $10^{10.3}M_{\odot}\leq M_{H}\leq10^{10.4}M_{\odot}$ to $10^{13.3}M_{\odot}\leq M_{H}\leq10^{13.6}M_{\odot}$ 
and concluded for each mass interval that halo accretion rates do not depend on environment at this redshift. 
We suggest that this apparent lack of environment dependence arises 
because the halos in the \cite{Percival} study and to a lesser extent the halos 
considered in the $z\sim2$ panel in
Fig.\ref{xi_percival_plot}, represent some of the most massive objects at $z\sim2$
and hence have bias factors $b>1$ \citep{Sheth99}. These structures are 
located at the highest peaks in the density field and so 
by computing the clustering amplitude of these objects
one is essentially measuring the clustering pattern of the highest density peaks at this redshift.
It is therefore unlikely that the highest mass halos experiencing different instantaneous accretion rates
differ in their clustering.
By contrast, the lower mass halos and subhalos in the $z<1$ panels are less biased and so more closely 
track the clustering of the underlying mass distribution. 

\subsection{Comparing dark halo growth with black hole growth}

Under the assumption that, on average, black hole growth 
traces dark halo growth (so-called ``pure coeval evolution''), 
M06 tested the predictions of equation (\ref{eMill}) 
for the evolution of the integrated AGN luminosity density for $z\leq3$. 
The coeval evolution model tests the hypothesis that the fractional mass
accretion rate onto black holes and onto halos are equal (i.e. $\dot{M}/M$ is the
same for both black holes and halos), and is
consistent with the tight relation inferred between black hole mass and galaxy bulge mass
\citep[but see \citealt{Batcheldor10} for an alternative interpretation]{Tremaine02}, 
and is easy to test.
M06 found the predicted integrated AGN luminosity density to be in remarkable 
agreement with the bolometric AGN luminosity density measured using hard X-ray data. 
They also found that for $z>0.5$ average black hole growth is well approximated by 
pure coeval evolution, but for $z<0.5$ the black hole luminosity density tails off 
more quickly than dark halo growth, and by $z=0$ is lower by a factor of $\sim 2$. 
They suggested that this slowdown in black hole accretion could be related to cosmic downsizing
\citep[e.g.][]{Barger}.

Their predictions for dark halo growth were, however, based on EPS theory. 
The simulation trajectories in Fig.\ref{halo_accn} 
show that EPS underestimates halo accretion for $z<1$, 
and at $z=0$ is a factor of $\sim1.5-2$ lower for all halo masses. 
This implies that present day dark halos could be accreting 
at fractional rates that are up to $\sim3-4$ times higher than their associated black holes. 
However, for $1<z<3$, the simulated dark halo accretion trajectories in Fig.\ref{halo_accn} 
are reasonably well approximated by EPS. We therefore suggest 
the following scenario: for $1<z<3$ black holes grow coevally with their dark hosts 
but for $z<1$, the epoch of cluster formation, 
their growth significantly decouples from that of their hosts.

It is still plausible that this decoupling is linked to the inference that high mass black holes
preferentially ``turn off'' at low redshifts, leaving the remaining accretion activity dominated
by low mass black holes \citep{Heckman}. The cause of such 
downsizing is often assumed to be connected to the physics of the baryon
component.  Our study reinforces this assumption: if downsizing were a ``whole halo'' phenomenon
it would be manifest in our dark-matter-only simulation, and its absence in our results
confirms that we should seek an explanation in the baryons.

\subsection{Is halo accretion via minor-mergers and diffuse accretion
the cause of radio-mode feedback?}\label{Croton_section}

A number of authors have developed semi-analytic models of galaxy-formation that are tuned to
reproduce the galaxy luminosity function at low redshift 
(e.g. \citealt{Bower06}; C06; \citealt{DeLucia06}).  A key ingredient of these models is a
low level of feedback from black hole accretion that arises in all galaxies and which 
increases in importance towards low redshifts.  The feedback mechanism has still
not been identified: luminous, high accretion-rate AGN only form a small subset of the galaxy
population at low redshift and seem unlikely to provide the required feedback in all galaxies.
\citet{Bower06} required black holes to have relatively high accretion Eddington ratios,
which may be inconsistent with observations: it seems that the accretion and an associated
outflow need to be hidden from view in a so-called ``radio-mode''.  C06 have
assumed that such a mode could be fuelled by Bondi accretion from the hot gas phase of their
model, but the observational evidence for such a mechanism has not been demonstrated either.

The survey of \cite{Ho97} revealed that a high fraction, over $40\%$ of nearby
galaxies rising to $50\%-75\%$ of bulge systems, host low luminosity
AGN (LLAGN), with the majority of LLAGN accreting at highly
sub-Eddington rates in the range
$10^{-5}<L_{bol}/L_{Edd}<10^{-3}$. 
\citet{Ho05} argued that these are systems where accretion occurs
via a radiatively-inefficient advection-dominated accretion flow (ADAF).
The accretion flow puffs up the inner disk and material is advected towards the
black hole \citep{Narayan, Ho02, Ho08}, with 
outflow being channelled along kinetic-energy-dominated
jets \citep{Collin, Ho05, Ho08}.  This finding leads us to suggest that LLAGN, fuelled
by low accretion rate ADAFs, may provide the radio-mode feedback.

In our dark-matter-only study, the integrated minor-merger and diffuse 
halo accretion rate density curves in
Fig.\ref{dm_by_m} increase in importance towards the present day for all halo masses.
This qualitatively agrees with the cosmological evolution of the black hole radio-mode integrated
accretion signal found for each of the different semi-analytic models (\citealt{Bower06}; C06).
We suggest that the periods when galaxy halo growth is
dominated by low accretion rate minor-mergers and diffuse accretion events,
are mirrored by low
accretion rates onto their associated black holes, and that those in
turn produce the LLAGN that may be the radio-mode required for the
feedback models.  

The integrated accretion rate density onto black holes residing in galaxy-mass halos 
that are accreting diffusely and via minor-mergers at $z=0$ is also very
similar to the integrated accretion rate density onto black holes residing in similar sized halos
found by C06, who argue that radio-mode feedback is more effective in more massive systems.
Our estimate for the total black hole accretion rate density tests the hypothesis that for 
black holes with mass $M_{BH}$ residing in halos with mass $M_{H}$,

\begin{equation}\label{radio-mode-estimate}
\sum_{i}\dot{M}_{i,BH}(z=0)\sim\alpha\frac{M_{BH}}{M_{H}}\sum_{i}\dot{M}_{i,H}(z=0)
\end{equation}
where $\alpha$ describes the non-linearity in the black hole - dark halo mass relation
and the index $i$ sums over all galaxy-mass dark halos and all black holes residing
in these halos. Equation (\ref{radio-mode-estimate}) assumes that
black hole growth positively traces dark halo growth, on average
(recent claims by \citealt{Kormendy11}, however, argue that for
bulgeless galaxies there is no such correlation between black holes and
their dark hosts, but the interpretation of this as meaning
that there is no such relation for all galaxies has been
clearly refuted by \citealt{Volonteri11}.
In what follows we do not address the 
reliability of the assumption in equation (\ref{radio-mode-estimate})
but rather test its prediction for black hole growth).
\cite{Ferrarese02} found that $\alpha=1.65$ and 
that galaxy-mass halos with $M_{H}\sim10^{12}M_{\odot}$ 
have a black hole - dark halo mass ratio of $\sim
10^{-5}$. According to Fig.\ref{dm_by_m} these halos 
with $\delta M_{H}/M_{H}\leq0.02$
have a total accretion rate density of
$\sim 7.6\times10^{7}M_{\odot}$Gyr$^{-1}$Mpc$^{-3}$ at $z=0$, 
which when substituted into equation (\ref{radio-mode-estimate})
yields a total black hole accretion rate density of
$\sim10^{-5.9}M_{\odot}$yr$^{-1}$Mpc$^{-3}$. This
is very similar to the integrated accretion rate density of
$\sim10^{-5.8}M_{\odot}$yr$^{-1}$Mpc$^{-3}$ onto
supermassive black holes at $z=0$ reported by C06.

The $\delta$ parameter ($\equiv\delta M_{H}/M_{H}$) 
is a free parameter in our model, but
we have found that adopting the more classical progenitor mass
ratio, $\chi$, to distinguish between merger type 
yields almost identical results to Fig.\ref{dm_by_m}.
This provides confirmation that our $\delta$ cuts
are indeed capable of separating minor- and major- merger channels.
The $\delta$ parameter is therefore probably no more unconstrained than $\chi$.

We conclude that the low rates of accretion onto dark halos,
driven by minor-mergers and diffuse accretion, 
may provide an alternative explanation to that proposed by C06 for
the radio-mode feedback needed to reproduce
the observed galaxy luminosity function.  The low redshift feedback
phenomenon and its cosmological evolution may be driven by the
cosmological evolution of halo minor-mergers and diffuse accretion
rather than requiring
accretion out of a hot gas phase.

\section{Conclusions}\label{conclusions_section}

Outputs from one of the high resolution dark-matter-only 
Horizon Project simulations have been used to investigate the environment and redshift 
dependence of accretion onto both halos and subhalos. We have developed a method
that computes the combined merger- and diffuse- driven accretion onto halos and all levels of substructure and
find that: 

\begin{itemize}
\item Halo accretion rates vary less strongly with redshift than predicted
by the EPS formalism. This offset in gradient for each halo mass
curve implies that perhaps minor adjustment to the EPS formula is required. 
\item Comparison with an observational study of black hole growth leads us to suggest
that dark halos at $z=0$ could be accreting at fractional rates that are up to $3-4$ times higher than
their black holes.
\item Halo growth is driven by minor-mergers and diffuse accretion
at low redshift. These latter accretion modes have both the correct 
cosmological evolution and inferred integrated black hole accretion rate density at $z=0$ to drive
radio-mode feedback, which has been hypothesised in recent semi-analytic galaxy-formation 
models as the feedback required to reproduce the galaxy luminosity function at low redshift.
Radio-mode feedback may therefore be driven by dark halo minor-mergers and diffuse accretion, 
rather than accretion of hot gas onto black holes, as has been recently argued. 
\item The low redshift subhalo accretors in the simulation 
form a mass-selected subsample safely above the mass resolution limit and
mostly reside in the outer regions of their
host, with $\sim 70\%$ beyond their host's virial radius, 
and are probably not therefore being significantly stripped of mass. These subhalos
accrete at higher rates than halos, on average, at low redshifts. 
We demonstrate that this is due to their enhanced mutual clustering 
at small scales: there is no significant difference
between the halo and subhalo accretor distributions of $\Delta v/v_{c}$, where
$\Delta v$ represents the relative velocity between an accretor's
main father and one of its other progenitors, and $v_c$
is the accretor's circular velocity.
The very frequent interactions with other subhalos of the same host
drive the high subhalo accretion rates.
\item Accretion rates onto halos and subhalos depend only weakly on environment at 
cluster scales. For halos, it appears that
the increased interaction rates in group- and cluster- mass environments
are not sufficiently large enough to significantly overcome the large halo relative
velocities, resulting in only a modest net increase in accretion at cluster scales.
The subhalo accretors only reside in the densest environments and they
are likely to be accreting mostly from their nearby subhalo neighbours, rather than from their host. 
We further demonstrate that halos accrete independently of their environment at $z\sim2$, as has 
been found by other authors, but show that this behaviour results from examining 
the clustering of the most massive halos with large bias factors. 
When less massive halos below $M_{\star}$ at low redshift are considered, a weak dependence between
accretion rate and environment at cluster scales arises.
\end{itemize}

\section*{Acknowledgements}

We are grateful to the Horizon Project team
for providing the simulation outputs
and to the anonymous referee whose insightful comments
have helped improve the quality of this paper.
The research of JD is partly funded by Adrian Beecroft,
the Oxford Martin School and the STFC.
HT is grateful to the STFC for financial support. 

\def\refname{REFERENCES}
\bibliographystyle{mn2e.bst}
\bibliography{hsub_accn_new}

\bsp

\label{lastpage}
\end{document}